\documentclass[11pt]{article}
\usepackage{amsfonts}
\usepackage{amssymb}
\usepackage{amscd}
\usepackage{amsmath}
\usepackage{amsthm}
\usepackage{latexsym}

\catcode`@=11
\renewcommand{\section}{\@startsection{section}{1}{0pt}{\medskipamount}
{\medskipamount}{\large\bf}}
\catcode`@=12

\def\a{\alpha}

\def\de{\delta}
\def\e{\epsilon}

\def\th{\theta}
\def\vt{\vartheta}
\def\l{\lambda}

\def\L{\Lambda}

\newcommand{\R}{\mathbb R}

\newcommand{\N}{\mathbb N}

\newcommand{\Lie}{{\mathop{\mbox{Lie}}\nolimits\,}}
\def\N2{$N{=}2$}

\newcommand{\pa}{\mbox{$\partial$}}

\def\dt#1{{\buildrel {\hbox{\bf .}} \over {#1}}}  

\def\ad{{\dt{\alpha}}}
\def\bd{{\dt{\beta}}}

\def\sfrac#1#2{{\textstyle\frac#1#2}}

\def\tr{{\rm tr}}
\def\ask#1{\buildrel ? \over #1}

\def\be{\begin{equation}}
\def\ee{\end{equation}}
\def\bea{\begin{eqnarray}}
\def\eea{\end{eqnarray}}
\def\zeile{\nonumber\\[.5ex] }
\def\gl#1{(\ref{#1})}

\newcommand{\Bmunu}{B_{\mu\nu}}
\newcommand{\alf}{\alpha}
\newcommand{\bet}{{\beta}}

\newcommand{\eps}{\varepsilon}

\newcommand{\alfp}{\alpha^{\prime}}

\newcommand{\psipnu}{\psi^{\nu}_+}

\newcommand{\psimnu}{\psi^{\nu}_-}

\newcommand{\psiquer}{\overline{\psi}^{\mu}}

\newcommand{\psiobmu}{\psi^{\mu}}
\newcommand{\psiobnu}{\psi^{\nu}}
\newcommand{\epsb}{\overline{\eps}}
\newcommand{\+}{+\!\!\!+}

\newcommand{\Xobmu}{X^{\mu}}
\newcommand{\Xobnu}{X^{\nu}}
\newcommand{\dunalf}{\pa_{\alf}}

\newcommand{\dunbet}{\pa_{\bet}}

\newcommand{\rhoalf}{\rho^{\alf}}
\newcommand{\rhobet}{\rho^{\bet}}

\newcommand{\epsalfbet}{\eps^{\alf\bet}}

\newcommand{\pasenk}{\pa_{\+}}
\newcommand{\papar}{\pa_=}
\newcommand{\intxi}{\int_{\Sigma}\!d^2\xi\;}

\newcommand{\dSigma}{|_{\mbox{\tiny{$\pa\Sigma$}}}}
\newcommand{\Emunu}{E_{\mu\nu}}
\newcommand{\Enumu}{E_{\nu\mu}}
\newcommand{\Jmunu}{J_{\nu}^{\mu}}
\newcommand{\Gmunu}{G^{\mu\nu}}
\newcommand{\gmunu}{g_{\mu\nu}}
\newcommand{\thetamunu}{\theta^{\mu\nu}}

\newcommand{\ab}{\bar{a}}

\begin{document}
\begin{titlepage}
\begin{flushright}
hep-th/0012200\\
ITP--UH--42/00\\
\end{flushright}

\vskip 2.0cm

\begin{center}

{\Large\bf  Open N=2 Strings in a B-Field Background \\
\vskip 0.2cm
and Noncommutative Self-Dual Yang-Mills }

\vspace{14mm}

{\large Olaf Lechtenfeld, \ Alexander D. Popov~$^*$ \ and \ Bernd Spendig}
\\[5mm]
{\em Institut f\"ur Theoretische Physik  \\
Universit\"at Hannover \\
Appelstra\ss{}e 2, 30167 Hannover, Germany }\\
{Email: lechtenf, popov, spendig@itp.uni-hannover.de}

\end{center}

\vspace{2cm}

\begin{abstract}
\noindent
In the presence of $D$-branes, fermionic \N2 strings in 2+2 dimensions 
can be coupled to a K\"ahler NS-NS two-form $B$. We present the corresponding 
action which produces \N2 supersymmetric boundary conditions and discuss the 
Seiberg-Witten zero-slope limit. After recalling the constraints on the 
Chan-Paton gauge group, we demonstrate for $U(n)$ groups that the open \N2 
string with a nonzero $B$-field coincides on tree level with noncommutative 
self-dual Yang-Mills. Several misconceptions of hep-th/0011206 are corrected.
\end{abstract}

\vfill

\textwidth 6.5truein
\hrule width 5.cm
\vskip.1in

{\small
\noindent ${}^*$
On leave from Bogoliubov Laboratory of Theoretical Physics, JINR,
Dubna, Russia}

\end{titlepage}

\section{Introduction and results}
\noindent
A constant NS-NS two-form background modifies string dynamics nontrivially 
if $D$-branes are present~\cite{douglas,cheung,chu,schomerus,ardalan}. 
In particular, open strings ending on $n$~coincident $D$-branes see a
deformed space-time metric~$G_{\mu\nu}$ and acquire a noncommutativity
parameter~$\theta^{\mu\nu}$. The latter means that the $D$-brane world volume
carrying the $U(n)$ Yang-Mills fields becomes noncommutative.
In an $\a'{\to}0$ limit which keeps the above open-string parameters finite 
the string indeed reduces to noncommutative $U(n)$ gauge theory on the brane
\cite{seiberg}. A restriction to $SO(n)$ or $Sp(n)$ subgroups is nontrivial
but emerges via orientifold projection~\cite{BSST}.

In the present letter we apply this analysis to \N2 strings.
Twenty years ago it was discovered~\cite{OV} that the open \N2 fermionic
string at tree level is identical to self-dual Yang-Mills field theory
in $2{+}2$ dimensions. The complete absence of a massive physical spectrum
ties in with the vanishing of all string amplitudes beyond three-point
although the one-loop structure~\cite{CLN} seems to be anomalous~\cite{CS2}.

It is expected that switching on a constant $B$-field background renders
the self-dual gauge theory noncommutative.
However, it also interferes with the global world-sheet supersymmetry
of fermionic strings in the superconformal gauge~\cite{lind}.
We shall show that a boundary term must be added to the \N2 string action
to preserve its supersymmetries for $B_{\mu\nu}{\neq}0$.
In addition, it turns out that the two-form $\Bmunu dx^{\mu}{\wedge}dx^{\nu}$ 
must be K\"ahler.

In order to verify the identity of the open \N2 string in a constant
$B$-field background with noncommutative self-dual Yang-Mills theory,
we shall discuss the factorization of open-string trees in this context
and prove that the four-point amplitude of noncommutative $U(n)$ self-dual 
gauge theory vanishes in accord with the string result.
As is demonstrated for the example of~$U(2)$, the restriction of
$U(n)$ to $SU(n)$ is not admissible~\cite{AD,Mat,Arm,BSST}.
This explains the erroneous result of a recent paper~\cite{KMP} 
claiming inconsistency for non-abelian gauge groups.

\section{Open N=2 strings}
\noindent
The  critical \N2 string lives in 2+2 real or equivalently 1+1 complex 
dimensions. Put differently, the string world sheet $\Sigma$ is embedded into 
a four-dimensional target space with signature $(2,2).$ The Brink-Schwarz 
action~\cite{brink} for the \N2 string in $\R^{2,2}$ is given by\footnote{
We use $\rho^0=\left(\begin{matrix} 0 & -1 \\ 1 & 0 \end{matrix} \right),\; 
\rho^1=\left(\begin{matrix} 0 & 1 \\ 1 & 0 \end{matrix} \right),\; 
\{\rho^m,\rho^n\}=2\eta^{mn},\;(\eta^{mn})=
\left(\begin{matrix} -1 & 0 \\ 0 & 1 \end{matrix} \right), \;
\xi^0=\tau,\; \xi^1=\sigma.$ 
The space-time metric is $(\eta_{a\ab})=\zeta 
\left(\begin{matrix} 1 & 0 \\ 0 & -1 \end{matrix} \right),$ 
where $\zeta > 0$ is a real scaling parameter. 
For the definition of $D_{\alf},\overline{\psi},\overline{\chi}, e$ 
see~\cite{bischoff}. }
\begin{align}
S\ =\ -\frac{1}{2\pi\alfp}\intxi e\,\big\{\sfrac{1}{2}h^{\alf\bet}
\dunalf X^{-\ab}\dunbet X^{+a} + \sfrac{i}{2}\,\overline{\psi}^{-\ab}\rhoalf
{\stackrel{\leftrightarrow}{D}}_{\alf}\psi^{+a} + 
A_{\alf}\overline{\psi}^{-\ab}\rhoalf\psi^{+a} \phantom{xxxxxxxx} \nonumber \\ 
+\ (\dunalf X^{+a} + \overline{\chi}_{\alf}^{-}\psi^{+a})
\overline{\psi}^{-\ab} \rhobet\rhoalf\chi_{\bet}^+ + 
(\dunalf X^{-\ab} +\overline{\psi}^{-\ab}\chi_{\alf}^{+})
\overline{\chi}_{\bet}^-\rhoalf\rhobet\psi^{+a} \big\}\,\eta_{a\ab} \quad.
\end{align}
The matter fields $X^{+a}$ and $\psi^{+a}$ are complex valued 
($X^{-\ab}{=}(X^{+a})^*,\;\psi^{-\ab}{=}(\psi^{+a})^*$), 
so that the space-time indices $a,\ab=1,2$ run over two values only. 
The fields are coupled to the \N2 supergravity multiplet consisting of the 
zweibein $e_{\alf}^n$ (related to the world-sheet metric $h_{\alf\bet}$ via 
$h_{\alf\bet}=\eta_{mn}e_{\alf}^m e_{\bet}^n$), the complex gravitino 
$\chi_{\alf}^{\pm}$ and the $U(1)$ connection $A_{\alf}$. 
Using symmetries of the action (see e.g.~\cite{bischoff} for a discussion) 
one can locally gauge away all gravitational degrees of freedom. 
In this superconformal gauge the action becomes 
\begin{equation}\label{n2action}
S\ =\ -\frac{1}{4\pi\alfp}\intxi \eta^{\alf\bet}
\big(\pa_{\alf} X^{-\ab}\pa_{\bet} X^{+a} 
+\sfrac{i}{2} {\overline \psi}^{-\ab}\rho_{\alf} \pa_{\bet}\psi^{+a}
+\sfrac{i}{2} {\overline \psi}^{+a}\rho_{\alf} \pa_{\bet}\psi^{-\ab}\big)
\,\eta_{a\ab}\quad.
\end{equation}

We now switch to a real notation via $(\pm a) \rightarrow (\mu),$ 
with $ \mu,\nu,\ldots=1,2,3,4.$  To be more explicit,
\begin{align}
X^1\ :=\ \frac{1}{2}(X^{+1}+X^{-1})\quad ,&\hspace{1cm} 
X^2\ :=\ \frac{1}{2i}(X^{+1}-X^{-1})\quad , \nonumber \\
X^3\ :=\ \frac{1}{2}(X^{+2}+X^{-2})\quad ,&\hspace{1cm} 
X^4\ :=\ \frac{1}{2i}(X^{+2}-X^{-2}) \quad ,
\end{align}
and analogously for the fermionic fields. 
The action functional then takes the form 
\begin{align}\label{action}
S\ &=\ -\frac{1}{4\pi\alfp}\intxi  \eta^{\alf\bet}\big(
\dunalf \Xobmu \dunbet\Xobnu 
+ i\,\overline{\psi}^{\mu}\rho_{\alf}\dunbet\psi^{\nu}\big)\,\gmunu\quad ,
\end{align}
with $(\gmunu)=\zeta \;{\rm diag}(+1,+1,-1,-1).$ 
This action enjoys a residual gauge invariance under \N2 superconformal 
transformations. In particular, the rigid \N2 supersymmetry transformations 
have the form~\cite{zumino, alvafreedman}
\begin{align}
\delta\Xobmu\ &=\ \epsb_1\psi^{\mu}+\Jmunu\epsb_2\psi^{\nu}\quad, \nonumber \\
\delta\psi^{\mu}\ &=\ -i\rhoalf\dunalf\Xobmu\eps_1
+ i \Jmunu\rhoalf\dunalf\Xobnu\eps_2\quad. 
\end{align}
Here $(\Jmunu)$ is a constant complex structure compatible with our metric, 
i.e.~$\gmunu J_{\lambda}^{\nu}+J_{\mu}^{\nu}g_{\lambda\nu}=0$.

\section{N=2 supersymmetric boundary conditions}
\noindent
We now turn our attention to \N2 open strings in a $B$-field background. 
Since $B$-field components not parallel to a $D$-brane world volume 
can be gauged away, we shall consider $n$ concident $D3$-branes in order to 
allow for the most general $B$-field configuration. Let us investigate 
how a $B$-field can be coupled to supersymmetric $2d$ matter fields so that 
the action is still globally supersymmetric. 

The gauge-fixed action functional 
derived from the standard superfield action is
\begin{align}\label{actionB}
S\ =\ -\frac{1}{4\pi\alfp}\intxi 
\big(\eta^{\alf\bet}\gmunu+\epsalfbet 2\pi\alfp \Bmunu\big)
\big(\dunalf \Xobmu \dunbet\Xobnu + 
i\,\overline{\psi}^{\mu}\rho_{\alf}\dunbet\psi^{\nu}\big)\quad .
\end{align}
The boundary conditions for $\Xobmu$ following from this action read 
\begin{equation}\label{bosbound}
(\Enumu \pasenk X^{\nu}-\Emunu \papar \Xobnu )\dSigma\ =\ 0\quad ,
\end{equation}
where 
\begin{equation}
\Emunu\ :=\ \gmunu+2\pi\alfp \Bmunu \quad , 
\end{equation}
and $\pa \Sigma=\{\xi^1{=}0,\pi\},$ 
while $\pasenk=\pa_0{+}\pa_1$ and  $\papar=\pa_0{-}\pa_1.$
The boundary conditions for $\psi^{\mu}$ must get mapped to~(\ref{bosbound}) 
under supersymmetry. The appropriate fermionic boundary conditions are 
(see e.g.~\cite{seiberg, lind}) 
\begin{equation}\label{fermbound}
(\Enumu \psipnu-\gamma \Emunu \psimnu)\dSigma\ =\ 0\quad ,
\end{equation}
where we use the fact\footnote{
Recall that a Majorana spinor $\varphi$ (in 1+1 dimensions) has two components 
$\varphi^{\pm}=\frac{1}{2}(1\pm \rho^1\rho^0)\varphi.$ Furthermore, 
$\overline{\varphi}\equiv (\varphi_+,\varphi_-)=\varphi^{\dagger}\rho^0.$} 
that $\eps_i^-=\eps_i^+$ at $\sigma{=}0$ 
and $\eps_i^-=\gamma\eps_i^+$ at $\sigma{=}\pi$ 
($\gamma{=}{+}1$ for the Ramond sector, 
$\gamma{=}{-}1$ for the Neveu-Schwarz sector).
 
A straightforward calculation shows that the fermionic boundary conditions 
derived from~(\ref{actionB}) are {\em inconsistent\/} with~(\ref{fermbound}). 
It was shown in~\cite{lind} that the $N{=}1$ fermionic string requires 
adding two $B$-dependent boundary terms to restore supersymmetry. 
This leads to the following expression for the $N{=}1$ string action:
\begin{equation}\label{action2}
S\ =\ -\frac{1}{4\pi\alf^{\prime}}\!\intxi  \big[
(\eta^{\alf\bet}\gmunu+\epsalfbet 2\pi\alfp\Bmunu)\,\dunalf\Xobmu\dunbet\Xobnu 
+ i \Enumu\psiquer\rhoalf\dunalf\psi^{\nu}\big]\quad .
\end{equation} 
For the \N2 string we find the same result. 
Furthermore, the second supersymmetry applied to the action 
leads to additional equations,
\begin{equation}\label{fermsup}
(\Enumu J_{\lambda}^{\nu}\pasenk X^{\lambda}-
 \Emunu J_{\lambda}^{\nu}\papar X^{\lambda})\dSigma\ =\ 0\quad .
\end{equation}
These  conditions are equivalent to~(\ref{bosbound}) and thus 
pose no further constraint only if we demand that
\begin{align}
\gmunu J_{\lambda}^{\nu}+J_{\mu}^{\nu}g_{\lambda\nu}\ =\ 0 
\qquad\mbox{and}\qquad
\Bmunu J_{\lambda}^{\nu}-J_{\mu}^{\nu}B_{\lambda \nu}\ =\ 0\quad . 
\end{align}
These relations mean that $(\gmunu)$ is a hermitian metric and 
$\Bmunu dx^{\mu}{\wedge}dx^{\nu}$ has to be a K\"ahler two-form on $\R^{2,2},$ 
i.e.~a closed two-form compatible with the complex structure $J{=}(\Jmunu)$.
It is important to notice that the action functional~(\ref{action2}) 
cannot be written in terms of superfields. In particular, the action used 
in~\cite{KMP} is not \N2 supersymmetric without adding boundary terms.

\section{Seiberg-Witten limit}
\noindent
We now want to investigate the effects of background $B$-fields 
on open \N2 strings and exhibit their effective field theory. 
The starting point is the form of the open-string 
correlators~\cite{schomerus, seiberg},
\begin{align}\label{propa}
\langle \Xobmu(\tau)\,\Xobnu(\tau^{\prime})\rangle\ &=\ 
-\alfp \Gmunu\,\ln(\tau-\tau^{\prime})^2\ +\
\sfrac{i}{2}\thetamunu\,\eps(\tau-\tau^{\prime})\quad , \\
\langle \psiobmu(\tau)\,\psiobnu(\tau^{\prime})\rangle\ &=\ 
\frac{\Gmunu}{\tau-\tau^{\prime}}\quad ,
\end{align}
for $\tau,\tau^{\prime}\in \pa\Sigma.$ Here, 
$[E^{-1}]^{\mu\nu}\equiv[(g+2\pi\alf^{\prime}B)^{-1}]^{\mu\nu}=
\Gmunu+\frac{1}{2\pi\alfp}\theta^{\mu\nu}$ 
yields the effective metric $G_{\mu\nu}$ seen by the open string 
and gives rise to the noncommutativity parameter $\theta^{\mu\nu}$ 
appearing in $[\Xobmu(\tau),\Xobnu(\tau)]=i\thetamunu$~\cite{schomerus}. 
With an appropriate choice of the $SO(2,2)$ generators~\cite{ivanova}, 
the matrices $J$ and $B$ can be written in terms of the generators of a 
$U(1)\times U(1)$ subgroup of $SO(2,2).$
Then, the complex structure $J$ and the most general `magnetic' $B$-field 
are expressed as
\begin{align}\label{mat}
J= (\Jmunu)=\left(\begin{matrix} 
0 & 1 &0 &0 \\ -1 & 0 & 0 &0 \\ 0&0&0&1 \\ 0&0&-1&0 
\end{matrix} \right)\qquad{\rm and}\qquad
B=(\Bmunu)&=\left(\begin{matrix} 
0 & B_1 &0 &0 \\ -B_1 & 0 & 0 &0 \\ 0&0&0&B_2 \\ 0&0&-B_2&0 
\end{matrix} \right)\quad .
\end{align}
In this basis we obtain
\begin{align}
(G^{\mu\nu})\ &=\ \left(\begin{matrix} 
\frac{\zeta}{\zeta^2+(2\pi\alf^{\prime}B_1)^2} & 0 &0 &0 \\ 
0 & \frac{\zeta}{\zeta^2+(2\pi\alf^{\prime}B_1)^2} & 0 &0 \\ 
0&0&-\frac{\zeta}{\zeta^2+(2\pi\alf^{\prime}B_2)^2}&0 \\ 
0&0&0&-\frac{\zeta}{\zeta^2+(2\pi\alf^{\prime}B_2)^2} 
\end{matrix} \right) \quad ,
\end{align}
\begin{align}
(\theta^{\mu\nu})\ &=\ \left(\begin{matrix} 
0&-\frac{(2\pi \alfp)^2 B_1}{\zeta^2+(2\pi\alf^{\prime}B_1)^2}  &0 &0 \\ 
\frac{(2\pi \alfp)^2 B_1}{\zeta^2+(2\pi\alf^{\prime}B_1)^2} &0  & 0 &0 \\ 
0&0& 0& -\frac{(2\pi \alfp)^2 B_2}{\zeta^2+(2\pi\alf^{\prime}B_2)^2} \\ 
0&0&\frac{(2\pi \alfp)^2 B_2}{\zeta^2+(2\pi\alf^{\prime}B_2)^2} & 0 
\end{matrix} \right)\quad . 
\end{align}
Note that for $B_2{=}{-}B_1$ the background will be self-dual, 
and the action~(\ref{action2}) will have $N{=}4$ supersymmetry
\cite{alvafreedman}. 

We also introduce a tetrad $e_{\hat{\mu}}=(e_{\hat{\mu}}^{\nu})$ 
related to the metric $G$ by the formula
\begin{equation}\label{tetrad}
\Gmunu\ =\ e_{\hat{\sigma}}^{\mu}\,e_{\hat{\lambda}}^{\nu}\,
\eta^{\hat{\sigma}\hat{\lambda}} \quad ,
\end{equation} 
where $(\eta^{\hat{\mu}\hat{\nu}})={\rm diag}(+1,+1,-1,-1)$ is the metric in 
the orthonormal frame and $\hat{\mu},\hat{\nu}=1,\ldots,4$ are Lorentz indices.

Next we calculate the effective open-string coupling $G_s$ 
which is related to the closed-string coupling $g_s$ via 
$G_s=g_s [{\rm det} G/{\rm det}(g+2 \pi \alfp B)]^{1/2}$ and obtain
\begin{equation}
G_s\ =\ g_s\Bigl[\bigl(1+(2\pi\frac{\alf^{\prime}}{\zeta}B_1)^2\bigr)
\bigl(1+(2\pi\frac{\alf^{\prime}}{\zeta}B_2)^2\bigr)\Bigr]^{1/2}\quad .
\end{equation}

The Seiberg-Witten limit consists of taking $\a'{\to}0$ while sending 
$\zeta \sim (\alfp)^2 \to 0$ (and therefore $\gmunu \to 0$) 
so that $G,\; G^{-1}$, and $\theta$ remain finite. 
This $\alfp \sim \zeta^{1/2} \to 0$ limit is equivalent to the limit 
$B \to \infty$~\cite{seiberg}.
We arrive at the following effective open-string coupling, 
\begin{equation}\label{Gs}
G_s\ \rightarrow\ \frac{g_{{\rm YM}}^2}{2\pi}\ \equiv\ 
4\pi^2\,|B_1B_2|\ =\ const\quad,
\end{equation}
since $g_s \sim \zeta \sim (\alfp)^2.$ 
The inverse open string metric $(\Gmunu)$ and the matrix $(\thetamunu)$ become
\begin{equation}
(\Gmunu)\rightarrow \left(\begin{matrix} 
\frac{1}{(2\pi B_1)^2} & 0 &0 &0 \\ 0& \frac{1}{(2\pi B_1)^2} &  0 &0 \\ 
0&0& \frac{-1}{(2\pi B_2)^2}&0 \\ 0&0&0& \frac{-1}{(2\pi B_2)^2} 
\end{matrix} \right)\quad{\rm and}\quad
(\thetamunu)\rightarrow \left(\begin{matrix} 
0&-\frac{1}{B_1}  &0 &0 \\ \frac{1}{B_1} &0  & 0 &0 \\ 
0&0& 0& -\frac{1}{B_2} \\ 0&0&\frac{1}{B_2} & 0 
\end{matrix} \right)\; .
\end{equation}

A remark is in order. Kumar et al.~\cite{KMP} use a notation similar to ours. 
Their choice of $B$-field, though, leads to the `electric' type of two-form, 
i.e.~their $B$-field has non-vanishing components simultaneously in space and 
time direction. In~\cite{susskind} this type of field has been considered in 
much detail, and it was shown that it does not admit a zero-slope limit 
which produces a field theory on a noncommutative space-time. 
It is not clear to us how a $*$ product and noncommutative gauge field theory 
can appear in~\cite{KMP} without the $\alfp \rightarrow 0$ limit.

\section{Factorization of open-string trees}
\noindent
In the absence of a $B$-field background ($\th{=}0$)
it is well known~\cite{GSW} that the factorization properties of open string
amplitudes (as required for unitarity) restrict the possible Chan-Paton
gauge groups to $U(n)$, $SO(n)$, and $Sp(n)$. 
In addition, one observes~\cite{Pol} that the $U(1)$ part of $U(n)$ decouples
from all amplitudes; hence, $SU(n)$ is admissible as well.

It is natural to ask whether turning on a non-vanishing constant $B$-field 
background further constrains the set of allowed Chan-Paton labels.
This question has been answered in the appendix of~\cite{AD} 
(correctly for $U(n)$) and in~\cite{BSST} (for $SO(n)$ and $Sp(n)$):
all these classical Lie groups are still allowed. However, the analysis
of~\cite{AD} immediately shows and ref.~\cite{BSST} explains that the 
restriction $U(n)\to SU(n)$ is no longer permitted because the $U(1)$ degree 
of freedom ceases to decouple. Let us briefly review the argument for $U(n)$.

The full $M$-particle open-string tree-level scattering amplitude reads
\begin{equation}\label{fullamp}
T(1,2,\ldots,M)\ =\ A(1,2,\ldots,M)\,\tr(\l_1\l_2\ldots\l_M)\,E(1,2,\ldots,M)
\ +\ \textrm{non-cyclic permutations}\quad ,
\end{equation}
where $A(1,2,\ldots,M)$ denotes the uncharged $\th{=}0$ primitive amplitude 
obtained from the disk diagram with external leg ordering $(1,2,\ldots,M)$, 
and the anti-hermitian matrix~$\l_i\in u(n)$ describes the group quantum number
of the $i$th external particle, $i{=}1,2,\ldots,M$. 
The only effect of the $B$-field background consists in multiplying each 
primitive amplitude with a phase~\cite{seiberg},
\begin{equation}
E(1,2,\ldots,M)\ :=\ \prod_{1\le j<\ell\le M} 
e^{-{i\over2}k_{j\mu}\th^{\mu\nu} k_{\ell\nu}} \quad,
\end{equation}
which, due to momentum conservation, is cyclically invariant just like 
the two factors it multiplies.

Let us focus on factorization.
Whenever a partial sum of external momenta goes on-shell,
the amplitude~$T$ develops a pole whose residue should factorize into
the $T$ amplitudes for the two halves of the cut diagram.
For a given pole, a subset of the permutations in \gl{fullamp} contributes. 
Generically,
\begin{equation}\label{afactorize}
A(1,2,\ldots,M)\ \sim\ {1\over m^2-s}\,\sum_X 
A(1,2,\ldots,P,X)\,A(X,P{+}1,\ldots,M) \quad ,
\end{equation}
where $s=-(k_1{+}k_2{+}\ldots{+}k_P)^2$, and $X$ runs over all states
in the spectrum with mass~$m$.
Similarly, for $u(n)$ (but not for $su(n)$!) one has
\begin{equation}\label{vons}
\tr(\l_1\l_2\ldots\l_M)\ =\ -2\,\sum_x 
\tr(\l_1\l_2\ldots\l_P\l_x)\,\tr(\l_x\l_{P+1}\ldots\l_M) \quad ,
\end{equation}
where $x$ labels a basis of anti-hermitian $u(n)$ generators normalized
to $\tr(\l_x\l_y)=-\sfrac12\de_{xy}$.
Note that the product $\l_1\l_2\ldots\l_L\notin u(n)$ but lies in the
universal enveloping algebra.
Finally, momentum conservation yields the factorization
\begin{equation}
E(1,2,\ldots,M)\ =\ E(1,2,\ldots,P)\,E(P{+1},\ldots,M) \quad.
\end{equation}

Taken together, one sees that each term in~\gl{fullamp} factorizes
correctly by itself in case of a $U(n)$ Chan-Paton group.
Yet, $T$~amplitude factorization functions under a somewhat weaker requirement.
Since the primitive amplitudes~$A$ at different leg orderings are further
related by
\begin{equation}
A(L,\ldots,2,1)\ =\ (-1)^L\,A(1,2,\ldots,L)
\end{equation}
for $L$~massless external states, we may group the permutations
in~\gl{fullamp} in quartets. With the help of
\begin{equation}
E(L,\ldots,2,1)\ =\ E(1,2,\ldots,L)^* \quad,
\end{equation}
the generic combination
$$
(1,\ldots\!,P,P{+}1,\ldots\!,M)+(P,\ldots\!,1,P{+}1,\ldots\!,M)+
(1,\ldots\!,P,M,\ldots\!,P{+}1)+(P,\ldots,1,M,\ldots\!,P{+}1)\nonumber  
$$
produces a factor of $\tr[\L(1,\ldots,P)\,\L(P{+}1,\ldots,M)]$ with
\begin{equation}
\L(1,2,\ldots,L)\ :=\ 
\l_1\l_2\ldots\l_L\,E(1,2,\ldots,L)\ -\
\l_L\ldots\l_2\l_1\,E(1,2,\ldots,L)^* \ \in\, u(n) \quad,
\end{equation}
multiplying the right-hand side of~\gl{afactorize}.
Hence, a subgroup $\mathcal{G}\subset U(n)$ is compatible with factorization,
if $\l_i\in\Lie(\mathcal{G})$ implies $\L(1,2,\ldots,L)\in\Lie(\mathcal{G})$ 
so that the trace may be split by inserting a complete basis 
$\{\l_x\}$ for $\Lie(\mathcal{G})$.

When $\th{=}0$, one has $E{=}1$, and the condition above holds for
the classical groups $U(n)$, $SO(n)$, and $Sp(n)$. In a $B$-field
background the condition becomes nontrivial already at $L{=}2$,
\begin{equation}
\l_1,\l_2\ \in \Lie(\mathcal{G}) \qquad \ask{\Longrightarrow} \qquad 
\L(1,2)\ \equiv\
\l_1\l_2\,E(1,2)\,-\,\l_2\l_1\,E(1,2)^*\ \in \Lie(\mathcal{G}) \quad.
\end{equation}
This seems to exclude $SO(n)$ and $Sp(n)$ groups~\cite{AD}.
However, a refined analysis employing the orientifold construction for
non-oriented open strings leads to a modified factorization condition,
which is indeed fulfilled by $SO(n)$ and $Sp(n)$~\cite{BSST}.
In contrast, a reduction of $U(n)$ to $SU(n)$ in the orientable case
is no longer possible because already $\tr[\L(1,2)]\neq0$,
indicating the fusion of two $SU(n)$-charged states to a $U(1)$-charged one. 

As mentioned before, the Seiberg-Witten limit ($\a'\to0$ but keeping
the open-string parameters finite) reduces the string to a noncommutative
quantum field theory. It is therefore not surprising that
the list of admissible open-string gauge groups for $\th{\neq}0$ 
is in perfect agreement with the list of possible noncommutative
Yang-Mills theories. In particular, the failure of the Moyal commutator
$f*g-g*f$ to close in $su(n)$ signals the necessity for the coupling
of an additional $U(1)$ gauge boson enlarging $SU(n)$ to $U(n)$~\cite{Mat,Arm}.

\section{Noncommutative self-dual Yang-Mills}
\noindent
We have already stated that beyond three-point all tree-level amplitudes 
of the \N2 fermionic string are known to vanish.
It is not always appreciated~\cite{KMP} that {\em complete\/} \N2 string 
amplitudes even at tree level include a sum over world-sheet instanton sectors 
labeled by the first Chern number of the gauged R-symmetry $U(1)$ bundle,
which turns each primitive amplitude into a function $A(G_s,\vt)$
not only of the open-string coupling~$G_s$ 
but also of an instanton (theta) angle~$\vt$~\cite{LS}.
Surprisingly, $SO(2,2)$ `Lorentz' transformations treat 
$\sqrt{G_s}(\cos{\vt\over2},\sin{\vt\over2})$ as a $(\sfrac12,0)$ spinor, so 
that we may put $G_s{=}1$ and $\vt{=}0$ in a suitable Lorentz frame~\cite{LS}.
The resulting three-string amplitude\footnote{
$k_i^+{\wedge} k_j^+:=k_{i4}k_{j1}-k_{i4}k_{j3}+k_{i2}k_{j1}-k_{i2}k_{j3}-
(i\leftrightarrow j)$} 
(in flat $\R^{2,2}$ with $B{=}0$),
\begin{equation}\label{cubicL}
T_3(1,2,3)\ =\ A_3(1,2,3)\,\tr(\l_1\l_2\l_3)+A_3(2,1,3)\,\tr(\l_2\l_1\l_3)\ =\
k_1^+{\wedge} k_2^+\,\tr(\l_{[1}\l_{2]}\l_3) \quad,
\end{equation}
represents the totally symmetric cubic interaction 
of the Leznov~\cite{L} prepotential~$\phi$~ 
for self-dual Yang-Mills theory~\cite{LS}. 
For more than three external legs, any tree-level \N2 string scattering
vanishes already on the level of the {\em primitive\/} amplitudes,
$A(1,2,\ldots,L{>}3)=0$, thanks to the kinematical identity
\begin{equation}\label{magic}
k_1^+{\wedge} k_2^+\,{1\over s_{12}}\,k_3^+{\wedge} k_4^+\ +\
k_2^+{\wedge} k_3^+\,{1\over s_{23}}\,k_1^+{\wedge} k_4^+\ +\
k_3^+{\wedge} k_1^+\,{1\over s_{31}}\,k_2^+{\wedge} k_4^+\ =\ 0
\end{equation}
valid only in $2{+}2$ dimensions~\cite{OV,hipp}.
It is very useful to note that this identity renders
\begin{equation}
\bar{A}_4(1,2,3,4)\ :=\ k_1^+{\wedge}k_2^+\,{1\over s_{12}}\,k_3^+{\wedge}k_4^+
\end{equation}
{\em totally antisymmetric\/} in all labels.

The vanishing of amplitudes implies the existence of symmetries and vice versa.
For the \N2 string an infinite number of tree-level scattering amplitudes 
vanishes and therefore an infinite number of symmetries is to be expected. 
For open \N2 strings these symmetries have been described in~\cite{ivle, lepo}. 

The more commonly used Yang gauge~\cite{Y} may also be obtained, 
by restricting oneself to the zero-instanton sector 
(or, equivalently, by averaging over~$\vt$).
Full \N2 string amplitudes, however, produce self-dual Yang-Mills in the
{\em Leznov gauge\/}.
The latter is also preferred by the simplicity of a merely quadratic field
equation, leading to no further field-theory vertices beyond the cubic one 
corresponding to~\gl{cubicL}.
Indeed, using again the `magical' identity~\gl{magic}
it is easy to prove that the sum of the $s$-, $t$-, and $u$-channel diagram 
for the field-theory four-point function already vanishes, leaving no room
for a quartic vertex~\cite{Par}. This result has been extended to all tree
amplitudes~\cite{marquart}.

The generalization to a non-vanishing constant $B$-field background is
straightforward. We switch on only `magnetic' components of the
$B$~field (see section 4), in order to allow for a Seiberg-Witten limit 
to noncommutative gauge theory.
The three-string (Leznov) amplitude is modified to
\begin{equation}\label{cubicnc}
T_3(1,2,3)\ =\ A_3(1,2,3)\ 
[\tr(\l_1\l_2\l_3) E(1,2,3) - \tr(\l_2\l_1\l_3) E(2,1,3) ] \quad , 
\end{equation}
while the dressing of the primitive amplitudes~$A$ by phase factors~$E$
does not alter their vanishing. These amplitudes lead to the cubic Lagrangian
\begin{equation}\label{lag}
\mathcal{L}\ =\ \frac{1}{2}\Gmunu\,\tr\,\pa_{\mu}\phi * \pa_{\nu} \phi \ +\ 
\frac{1}{3}\e^{\ad\bd}\,\tr\,\phi*\hat{\pa}_{0\ad}\phi*\hat{\pa}_{0\bd}\phi
\end{equation}
with noncommutative $*$ product. Here, 
$\hat{\pa}_{0\dot{0}}:=\hat{\pa}_{\hat{2}}{+}\hat{\pa}_{\hat{4}}$ and 
$\hat{\pa}_{0\dot{1}}:=\hat{\pa}_{\hat{1}}{-}\hat{\pa}_{\hat{3}},$ 
where $\hat{\pa}_{\hat{\mu}}:=e_{\hat{\mu}}^{\nu} \pa_{\nu}$ is defined 
with the help of the tetrad~(\ref{tetrad}).

In accord with the general discussion~\cite{seiberg} we expect the \N2 string
in a constant $B$-field background to be identical to noncommutative
self-dual Yang-Mills theory~\cite{taka, legare} in the Leznov gauge, 
as described by the Lagrangian~(\ref{lag}). Since the latter 
has only a cubic interaction vertex (for the Leznov prepotential~$\phi$),
all tree-level field-theory amplitudes (with more than three external legs) 
obtained using the Feynman rules based on~\gl{cubicnc} should be zero. 
As a nontrivial check, we should be able to reproduce the vanishing of 
the four-point function for noncommutative $U(n)$ self-dual Yang-Mills.

The {\em field-theory\/} four-point function~$T^{\rm Leznov}_4$ 
for the Leznov prepotential~$\phi\in u(n)$
is a sum over 24 permutations of
\bea
T_3(1,2,\cdot)\,{-1\over s}\,T_3(\cdot,3,4) 
&=& {-1\over s} A_3(1,2,\cdot)\,A_3(\cdot,3,4)\,
\sum_x \tr[\L(1,2)\l_x] \tr[\l_x\L(3,4)] \zeile 
&=& \sfrac12 \bar{A}_4(1,2,3,4)\,\tr[\L(1,2)\L(3,4)] \quad ,
\eea
where the last equation makes use of~\gl{vons}.
Due to its total antisymmetry $\bar{A}_4$ may be pulled out 
of the permutation sum, which reduces to
\begin{equation}
\sum_{\pi\in\mathcal{S}_4} (-)^\pi\,
\tr[\L(\pi_1,\pi_2)\L(\pi_3,\pi_4)]\ =\
4 \sum_{\pi\in\mathcal{S}_4} (-)^\pi\,
\tr[\l_{\pi_1}\l_{\pi_2}\l_{\pi_3}\l_{\pi_4}]\,E(\pi_1,\pi_2,\pi_3,\pi_4)
\quad.
\end{equation}
However, since each term under the sum is cyclically invariant,
the four contributions to any cycle cancel each other in pairs, 
leaving us with
\begin{equation}
T^{\rm Leznov}_4(1,2,3,4)\ =\ 0 \quad.
\end{equation}
Of course, the Yang gauge produces the same result.
Because the gradient of~$\phi$ yields the Yang-Mills gauge potential,
the same-helicity four-gluon amplitude emerges from $T^{\rm Leznov}_4$ 
by multiplication of leg factors and thus continues to vanish in the
noncommutative case.

\section{Four-point amplitude: an example}
\noindent
It has recently been claimed~\cite{KMP} that noncommutative self-dual
Yang-Mills (in the Yang gauge) descends from the \N2 string only for
{\em abelian\/} gauge groups, because the field-theory four-point function
(for the Yang prepotential) allegedly fails to vanish otherwise.
Although we have demonstrated in generality that proper factorization 
guarantees the agreement of string with field-theory amplitudes,
let us elucidate the error of~\cite{KMP} for the simplest non-abelian
gauge group admitted, $\mathcal{G}=U(2)$.
In this case, the Leznov prepotential consists of an $su(2)$ triplet~$\phi^T$
plus an $su(2)$ singlet~$\phi^S$ stemming from the $U(1)$ gauge boson.
As generators we take $\l_a={i\over2}\sigma_a$, $a{=}1,2,3$, and
$\l_0={i\over2}{\bf1}$.

The vertices involving triplet states are
\begin{equation}
T_3^{TTT}(1,2,3)\ =\ \sfrac12\,k_1^+{\wedge} k_2^+\, c_{12}\,\e_{123}
\qquad\textrm{and}\qquad
T_3^{TTS}(1,2,3)\ =\ -\sfrac12\,k_1^+{\wedge} k_2^+\, s_{12}\,\de_{12} \quad ,
\end{equation}
where 
\begin{equation}
c_{ij}\ :=\ \cos(\sfrac12 k_i\th k_j)
\qquad\textrm{and}\qquad
s_{ij}\ :=\ \sin(\sfrac12 k_i\th k_j) \quad.
\end{equation}

Let us compose the four-triplet amplitude~$T^{\rm Leznov}_4$.
Triplet exchange in $s$-, $t$-, and $u$-channel yields
\bea
T_4^{(T)} &=& -\sfrac14\,\bar{A}_4(1,2,3,4) \sum_{x=1,2,3} [
c_{12}c_{34}\,\e_{12x}\e_{x34} +
c_{23}c_{14}\,\e_{23x}\e_{x14} +
c_{31}c_{24}\,\e_{31x}\e_{x24} ] \\[.5ex]
&=& -\sfrac14\,\bar{A}_4(1,2,3,4)\, \left[
\de_{12}\de_{34} (c_{23}c_{14}{-}c_{31}c_{24}) +
\de_{23}\de_{14} (c_{31}c_{24}{-}c_{12}c_{34}) +
\de_{31}\de_{24} (c_{12}c_{34}{-}c_{23}c_{14}) \right] \  ,
\nonumber
\eea
while singlet exchange produces
\begin{equation} \!\!\!\!\!\!\!\!\!
T_4^{(S)}\ =\ -\sfrac14\,\bar{A}_4(1,2,3,4)\, [
s_{12}s_{34}\,\de_{12}\de_{34} +
s_{23}s_{14}\,\de_{23}\de_{14} +
s_{31}s_{24}\,\de_{31}\de_{24} ] \quad. 
\phantom{xxxxxxxxxxxxxx}
\end{equation}
Even though the partial amplitudes do not vanish (for $\th{\neq}0$),
their sum $T^{\rm Leznov}_4=T_4^{(T)}+T_4^{(S)}$ does,
as may be verified from
\begin{equation}
c_{23}c_{14} - c_{31}c_{24} + s_{12}s_{34}\ =\ 0
\end{equation}
by employing momentum conservation.

For $\mathcal{G}{=}U(n)$, $\e_{abc}\to f_{abc}$,
but additional cubic couplings appear due to the
non-vanishing of the symmetric $SU(n)$ rank-three tensor~$d_{abc}$.
In conclusion, noncommutative self-dual Yang-Mills (at tree level)
is identical to the \N2 string in a constant $B$-field, as long
as one does not attempt to use pure $SU(n)$ or an exceptional gauge group.

\vskip2.cm

\noindent
{\bf\large Acknowledgements}\\[6pt]
This work was partially supported by the German Science Foundation (DFG)
under the grant LE 838/7-1 and the DFG-RFBR grant no. 99-02-04022.  
B.S.\ thanks the Studienstiftung des deutschen Volkes for support.

\end{document}